\documentclass[twocolumn,aps,pre,showpacs,epsfig]{revtex4}
\usepackage[dvips]{graphics}
\usepackage{graphicx}
\usepackage{amsfonts}
\usepackage{amssymb}
\usepackage{amsmath}
\begin{document}
\title{Discrete breathers in BEC with two- and three-body
interactions in optical lattice}
\author{F.Kh. Abdullaev$^{1}$ , A. Bouketir$^{2}$, A. Messikh$^{3}$,
B.A. Umarov$^{1,3}$\footnote[7]{Corresponding author
(b\_umarov@yahoo.com)}} \affiliation{$^{1}$Physical-Technical
Institute, Uzbek Academy of Sciences, 700084, Tashkent-84,
G.Mavlyanov str, 2-b, Uzbekistan}
\affiliation{$^{2}$Department of Science in
 Engineering,  Faculty of Engineering}
\affiliation{$^{3}$Department of
Computational and Theoretical Sciences,
 Faculty of Science,  International Islamic University Malaysia, P.O.
Box 10, 50728 Kuala Lumpur, Malaysia}
\date{\today}
\begin{abstract}
We investigate the properties of discrete breathers in a
Bose-Einstein condensate with three-body interactions in optical
lattice. In the tight-binding approximation the Gross-Pitaevskii
equation with periodic potential for the condensate wavefunction
is reduced to the cubic-quintic discrete nonlinear Schr\"odinger
equation. We analyze the regions of modulational instabilities of
nonlinear plane waves. This result is important to obtain  the
conditions for generation of discrete solitons(breathers) in
optical lattice. Also using the Page approach, we derive the
conditions for the existence and stability for  bright discrete
breather solutions. It is shown that the quintic nonlinearity
brings qualitatively new conditions for stability of strongly
localized modes. The numerical simulations conform with the
analytical predictions.
\end{abstract}
 \maketitle
\section{Introduction}
Discrete breathers are important eigenmodes of the nonlinear
discrete and periodic systems and their existence has been
predicted for such systems as optical waveguide arrays, array of
Josephson junctions, magnetic chains etc \cite{rev,rev_f,Leder01}.
Recently discrete breathers in array of Bose-Einstein condensates
has attracted a great deal of attention
\cite{Tromb01,Abd01a,Kon04,Alfimov,Kevr04,AK,Morsh}. The localized
atomic matter wave in BEC in the optical lattice(a gap soliton)
has been observed recently in \cite{Ober04}. The theoretical
approaches mainly based on the analysis of the Gross-Pitaevskii
equation with periodic potential and cubic mean field
nonlinearity. It corresponds to taking into account elastic
two-body interactions and the potential of optical lattice. The
progress  with Bose-Einstein condensates on the surface of atomic
chips and in atomic waveguides involves strong compression of  BEC
and so to essential increasing the density of BECs. Then it  pose
the problem of the taking into account three-body interactions
effects. This interaction represent interest also for the
understanding of the fundamental limits for the functioning of
devices using BEC \cite{Pu}. The existence of three-body
interactions can play important  role in the condensate stability
\cite{Abd01b,Akhmediev}. Recently 3-body interaction arising due
to the Efimov resonance has been observed in an ultra-cold gas of
cesium atoms \cite{Kraemer}.

Thus represent interest to investigate properties of discrete
breathers in BEC in optical lattices when two- and three-body
interactions are taken into account.  One of the interesting
properties of this system is the existence of the gap-Townes
soliton\cite{AS05}. In the tight-binding approximation(see a
section below) we show that the GP equation with periodic
potential can be  reduced to the cubic-quintic discrete nonlinear
Schr\"odinger(CQDNLS) equation. This regime occurs for the deep
optical lattice, which is realized in typical experiments in the
BEC.

For the generation of discrete breathers the modulational
instability of nonlinear plane wave represent the possible
mechanism. Thus in this paper in the section 3 we discuss the
modulational instability in the CQDNLS equation, the stability
region of the wave in the lattice. In section 4 we discuss the
localized modes of the different symmetries and the conditions of
their stability.

\section{The model}

Let consider BEC with two and three body interactions in optical
lattice and in highly anisotropic trap corresponding to quasi 1D
geometry. The Gross-Pitaevskii equation is
\begin{widetext}
\begin{equation}
i\hbar\psi_{t} = -\frac{\hbar^2}{2m}\nabla^{2}\psi + \frac{m}{2}
(\omega_{\perp}^2 \rho^2 + \omega_{x}^{2}x^{2}) + V_{opt}(\rho,x)
- g_{1}|\psi|^{2}\psi - g_{2}|\psi|^{4}\psi,
\end{equation}
\end{widetext}
where $g_1, \, g_2$ are coefficients proportional to  two- and
three-body elastic interactions. For rubidium atoms it is expected
that the three-body interaction is attractive and $|g_3| \sim
10^{-26}-10^{-27}$cm$^6$/s \cite{Pu}. Here $\omega_{\perp} \gg
\omega_{x}$, $V_{opt} = V_{0}\sin^{2}(kx)R(\rho).$ We look for the
solution of the form
\begin{equation}
\psi(r,t) =\phi_{0}(\rho) \phi (x,t),
\end{equation}
where $\phi_{0}$ is the solution of the radial linear equation
\begin{equation}
\frac{\hbar^2}{2m}\nabla_{\rho}^{2}\phi_{0} + \frac{m}{2}
\omega_{\perp}^{2}\rho^{2}\phi_{0} = \frac{1}{2}\hbar
\omega_{\perp}\phi_{0}.
\end{equation}
The solution for the ground state is
$$\phi_{0} = \sqrt{\frac{1}{\pi a_{\perp}^2}}
\exp(-\frac{\rho^{2}}{2a_{\perp}^2}).$$

Multiplying both sides of the GP  equation by $\phi_{0}$ and
integrating over transverse variable $\rho$ we obtain the quasi 1D
GP equation
\begin{widetext}
\begin{equation}
i\hbar \phi_{t} =-\frac{\hbar^2}{2m}\phi_{xx} + (\frac{m}{2}
\omega_{x}^{2}x^{2} + V_{0}\sin^{2}(kx))\phi - \frac{g_{1}}{2\pi
a_{\perp}^2}|\phi|^{2}\phi - \frac{g_{2}}{3\pi^2
a_{\perp}^4}|\phi|^{4}\phi.
\end{equation}
\end{widetext}
Introducing the dimensionless variables
$$t = t\nu, x = kx, \phi =
u\sqrt{\frac{2a_{s}\omega_{\perp}}{\nu}}, \nu =
\frac{E_{R}}{\hbar},$$ where $E_{R} = \hbar^2 k^2/2m$ is the
recoil energy,
 we obtain the equation
\begin{equation}\label{GP}
iu_{t} + \frac{1}{2}u_{xx} + \lambda |u|^{2}u + \bar{\lambda}
|u|^{4}u - V_{0} \cos(2x)u =0.
\end{equation}
where $\lambda = \pm 1$ and
$$\bar{\lambda} = \frac{g_{2}}{3\pi^2 \hbar \nu a_{\perp}^4}
\left( \frac{\nu}{2a_{s}\omega_{\perp}} \right)^{5/2}.$$
In the case of deep optical lattice ($U \gg E_{R}$ i.e. $\nu \gg
1$) the tight-binding approximation can be used
\cite{Tromb01,Abd01a,Kon04,Alfimov}.  If the potential $U(x)$
periodic, i.e. $U(x + l) =U(x)$, the eigenvalue problem is:
\begin{equation}
\frac{d^{2}\phi_{\alpha,q}}{dx^2} + U(x)\phi_{\alpha,q} =
E_{\alpha,q} \phi_{\alpha,q},
\end{equation}
where for the optical lattice potential $U(x) = V_{0}\cos(2x)$.
Here $\phi_{\alpha,q}$ is the Bloch function, $\alpha$ is the
index labelling the energy bands $E_{\alpha}(q)$. Periodicity of
the lattice admits the expansion of the energy
$$E_{\alpha,q+q_{0}} = E_{\alpha,q} = E_{\alpha,-q},\
q_{0}=\frac{2\pi}{l}.$$ Thus we can expand the energy in the
Fourier series
\begin{equation}
E_{\alpha,q} = \sum_{n}\hat{\omega}_{n,\alpha}e^{iqnl}, \
\hat{\omega}_{n,\alpha} = \frac{l}{2\pi}\int_{-q_{0}/2}^{q_{0}/2}
E_{q,\alpha}e^{-iqnl}dq.
\end{equation}
For deep lattice case  it is convenient to use the Wannier
functions\cite{Alfimov}:
\begin{equation}
w_{n,\alpha} = \sqrt{\frac{l}{2\pi}}\int_{-q_{0}/2}^{q_{0}/2}
\phi_{\alpha,q}(x)e^{-iqnl}dq,
\end{equation}
and
\begin{equation}
\phi_{\alpha,q} =
\sqrt{\frac{l}{2\pi}}\sum_{n}w_{n,\alpha}(x)e^{iqnl}.
\end{equation}
The main property of Wannier functions that they are strongly
localized in the bottoms of the potential and so are suitable for
the description of strongly localized modes in the periodic
potential. They form an  orthonormal and a complet set
\begin{eqnarray}
\int_{-\infty}^{\infty}w_{n,\alpha}^{\ast}w_{n',\alpha'}(x)dx =
\delta_{n,n'}\delta_{\alpha,\alpha'},\nonumber\\
\sum_{n,\alpha}w_{n,\alpha}^{\ast}(x)w_{n,\alpha}(x') =
\delta(x-x').
\end{eqnarray}
Let look  for the solution of CQ GP equation of the form of
expansion in Wannier functions
\begin{equation}
u(x,t) = \sum_{n,\alpha}c_{n,\alpha}(t)w_{n,\alpha}(x).
\end{equation}
Substituting this expansion into the GP equation (\ref{GP}) we
obtain the system of equation for coefficients $c_{n,\alpha}$

\begin{widetext}
\begin{eqnarray}
ic_{n,\alpha;t} &=& \sum
c_{n_{1},\alpha}\hat{\omega}_{n-n_{1},\alpha}
+ \lambda
\sum_{\alpha_{1},\alpha_{2},\alpha_{3}}\sum_{n_{1},n_{2},n_{3}}
c_{1}^{\ast}c_{2}c_{3}W^{nnnn}_{\alpha\alpha\alpha\alpha}
+
\bar{\lambda}\sum_{5\alpha}\sum_{5n}c^{\ast}_{1}c^{\ast}_{2}c_{3}c_{4}c_{5}
W_{6\alpha}^{6n},
\end{eqnarray}
\end{widetext}
where $$W_{nn_1n_2n_3}^{\alpha\alpha_1\alpha_2\alpha_3} =
\int_{-\infty}^{\infty}w_{n,\alpha}w_{n_1,\alpha_1}w_{n_2,\alpha_2}w_{n_3,\alpha_3}\
dx.$$ For deep optical lattice $V_0 > 5 E_R$ are important terms
with $n = n_{1}= n_{2} = n_{53}$ i.e. we are restricted by the
band $\alpha$. Then the system is reduced for equation

\begin{widetext}
\begin{equation}
iC_{n,\alpha;t} = \hat{\omega_{1,\alpha}}(C_{n-1,\alpha} +
C_{n-1,\alpha}) + \lambda W_{4\alpha}^{4n}|C_{n}|^{2}C_{n} +
\bar{\lambda} W_{6\alpha}^{6n} |C_{n}|^{4}C_{n}.
\end{equation}
\end{widetext}
By the transformation $v_{n}
=C_{n,\alpha}\sqrt{W_{4\alpha}^{4n}}C_{n}$ this equation can be
transformed into
\begin{equation}\label{dcq}
iv_{n,t} + \kappa(v_{n+1} + v_{n-1}) + \lambda |v_{n}|^{2}v_{n} +
{\gamma} |v_{n}|^{4}v_{n}=0.
\end{equation}
Here $\kappa = \hat{\omega_{1,\alpha}}, \gamma = \bar{\lambda}
W_{6\alpha}^{6n}/(2(W_{4\alpha}^{4n})^{2}).$ It is the
cubic-quintic discrete nonlinear Schr\"odinger(CQDNLS) equation.
As showed the analysis performed in \cite{Alfimov}, the region of
parameters exists, when the tight-binding approximation describes
the GP equation with periodic potential with a high accuracy. It
is achieved for the amplitudes of the periodic potential $|V_0| >
5 E_R, \,$ and $E_R$ is the recoil energy.  That corresponds to
the experimentally realized values. When $\gamma =0$ the problem
is reduced to considered previously by
\cite{Tromb01,Abd01a,Alfimov}. In the recent work \cite{Malomed06}
the case when $\lambda >0, \gamma < 0$, motivated by the analysis
of localized states in array of waveguides with saturable
nonlinearity, has been considered. In the BEC case the parameter
$\gamma$ can has any sign. The conserved quantities are: The
norm(number of atoms) $N$ is:
\begin{equation}\label{norm}
N = \sum_{n}|v_{n}|^{2},
\end{equation}
and the Hamiltonian $H$ is:
\begin{equation}\label{ham}
H = -\sum_{n=-\infty}^{\infty}\left[ \kappa(v_{n}^{\ast}v_{n+1} +
v_{n}v_{n+1}^{\ast})
 + \frac{\lambda}{2} |v_{n}|^{4} + \frac{\gamma}{3}
|v_{n}|^{6} \right].
\end{equation}
The equation of motion  is:
\begin{equation}
iu_{n,t} = \frac{\delta H}{\delta u_{n}^{\ast}}.
\end{equation}
The momentum is not conserved due to the discretness of the
system. Due to the periodicity of the system the Hamiltonian is
periodic with period $n=1$. The different excitations induces
different effective periodic potential like the Peierls-Nabarro
potential. The height of the barrier depends on the localization
of the discrete soliton solution - grows with the decreasing of
the soliton width. The analysis of the Hamiltonian can help for
the calculation of the PN barrier between different discrete
solitons and so help for the investigation of the stability of
nonlinear modes \cite{Cai}.

\section{Modulational instability of nonlinear plane
waves}

The nonlinear discrete equation has the plane wave solution
\begin{equation}\label{eq18}
v_n(t)=\nu_0 e^{i(qn+\omega t)},
\end{equation}
with the dispersion relation
\begin{equation}\label{eq19}
\omega = 2k\cos q +\lambda {\nu_{0}}^2 +\gamma {\nu_{0}}^{4}.
\end{equation}

For special importance here are the staggered $q = \pi$ and
unstaggered $q = 0$ solutions\cite{ADG} where the dispersion
relation can be rewritten as
\begin{equation}\label{eq20}
\omega = \pm 2k +\lambda {\nu_{0}}^2 +\gamma {\nu_{0}}^4.
\end{equation}
The linear stability of the solution (\ref{eq18}) and (\ref{eq19})
can be investigated by looking for a solution in the form
\begin{equation}\label{eq21}
v_n(t)=(\nu_0+\delta v_n (t)) e^{i(qn+\omega t)},
\end{equation}

 where $\delta v_n (t)$ is a small perturbation on the
carrier wave, substituting (\ref{eq21}) in (\ref{dcq}) and keeping
only the linear terms on $\delta v_n (t)$ and $\delta v^\ast _n
(t)$ we obtain the following equation

\begin{widetext}
\begin{equation}\label{eq22}
i\delta v_{n,t} +k\cos q (\delta v_{n+1}+\delta v_{n-1}-2 \delta
v_{n})+(\lambda+2\gamma {\nu_0}^2)(\delta v_{n}+\delta
v^\ast_{n})=0.
\end{equation}
\end{widetext}
Considering a modulation in the form
\begin{equation}
\delta v_{n} (t) = \left(
\begin{array}{cc}
\alpha  \\
\bar{\beta}
\end{array}\right) e^{i(Qn+\Omega t)},
\end{equation}
 where $Q$ and $\Omega$  are the wave number and the
frequency of the linear modulation waves respectively then
substituting it into (\ref{eq22}) we obtain the following system
of linear equations
\begin{eqnarray}
i \Omega \alpha +4k \cos q \sin^2 (Q/2)\bar{\beta}=0, \nonumber\\
-(4k\cos q \sin^2 (Q /2)+2(\lambda+2\gamma \nu_0 ^2)\nu_0
^2)\alpha +i\Omega \bar{\beta}=0.
\end{eqnarray}
 In order to have a
nontrivial solution for the system the determinant of the
coefficient matrix should be zero that leads to the following
dispersion law

\begin{widetext}
\begin{eqnarray}\label{eq25}
\Omega ^2 =8k \sin^2(Q/2)[2k\cos^2 (q)\sin^2 (Q/2)-\cos q
(\lambda+2\gamma \nu_0 ^2) \nu_0 ^2].
\end{eqnarray}
\end{widetext}
 The plane wave solution (\ref{eq18}) is modulationally unstable
provided that $\Omega^2$ is negative and the gain is
\begin{widetext}
\begin{eqnarray}\label{eq26}
G=2|\sin(Q/2)|\sqrt{2k\cos q(\lambda+2\gamma \nu_0 ^2)\nu_0
^2-2k\cos q \sin^2(Q/2))}.
\end{eqnarray}
\end{widetext}
From the gain equation we can see that the unstaggered solution
$(q=0)$ are unstable whenever $\lambda \nu_0 ^2 +2\gamma \nu_0^4
>0$ and the staggered solutions $(q=\pi)$ are unstable
whenever $\lambda \nu_0 ^2 +2\gamma \nu_0^4 < 0$ which leads to
the following cases:

i) $q=0$. If $\lambda>0$, $\gamma<0$ then we have MI of the
unstaggered solution for $|\gamma|<\lambda/2\nu_0 ^2$ and if
$\lambda<0$, $\gamma>0$ then we have MI when
$|\gamma|>\lambda/2\nu_0 ^2$.

ii) $q=\pi$. If $\lambda>0$, $\gamma<0$ then we have MI of the
staggered solution for $|\gamma|>\lambda/2\nu_0 ^2$ and if
$\lambda<0$, $\gamma>0$ then we have MI when
$|\gamma|<\lambda/2\nu_0 ^2$.

 iii) In contradiction with the continuum case which
is stable for the set of nonlinearities $\lambda<0$, $\gamma<0$
the staggered solution is modulationally instable regardless the
value of any parameter. However the unstaggered solution should
satisfy the inequality $\lambda \nu_0 ^2 +2\gamma \nu_0 ^4>2k$,
which is almost satisfied all the time for strongly localized
modes, in order to have MI.

In the general case where $q$ can take any value we can
distinguish two regions of MI as it is shown in Fig. 1 for
positive $k \cos q$.

\begin{figure}[here]
\includegraphics[height=8cm, width=8cm]{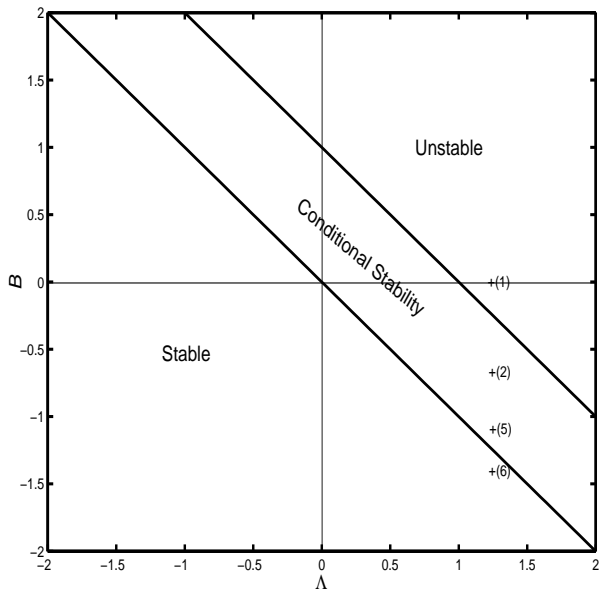}
\caption{Modulational instability regions as function of the
parameters $\Lambda=\gamma/2k{\nu_0}^2 \cos q$ and
$B=\beta/4{\nu_0}^2 \cos q$. with $\cos q>0$, the labelled points
are the points discussed in the text.  }
\end{figure}

One is fully unstable and the other is conditional means it
depends on the wave number $q$ of the carrier wave and the wave
number $Q$ of modulation wave as it is shown  in Fig. 2.

\begin{figure}[here]
\includegraphics[height=8cm, width=8cm]{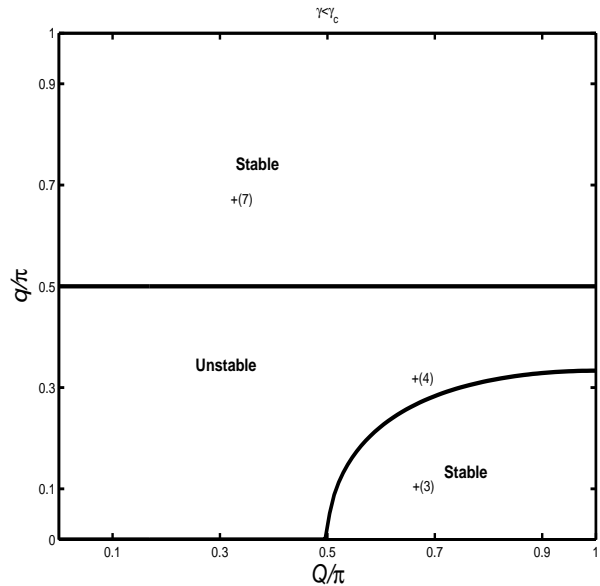}
\caption{Regions of modulational instability in the plane $(q,Q)$
for the conditional stability region in Fig. 1, the labelled
points are the points discussed in the text. }
\end{figure}

The conditional region represents the effect of the discreteness
on the modulational instability. Clearly we can see that if we fix
the cubic nonlinearity parameter and keep changing the quintic
nonlinearity parameter then we will have instability, conditional
instability and stability which shows that the quintic term can
lead to collapse of MI. The role of quintic term also is very
crucial on the wave numbers of the carrier and modulation waves
leading to MI, where increasing $\gamma$ in the negative direction
will shrink the MI region in $(q, Q)$ plane till a critical value
$\gamma_{cr}$. This eliminates all possible chances to get MI then
MI occurs again in the $(q, Q)$ plane in a mirror symmetry with
the one occured for $\gamma < \gamma_{cr}$. In the continuum
limit, when $Q<<1$ and $q<< 1$ equation (\ref{eq25}) reduced to
\begin{equation}\label{eq27}
\Omega^2=2kQ^2[kQ^2/2 -(\lambda+2\gamma \nu_0 ^2)\nu_0 ^2]
\end{equation}
and coincides with the one obtained in \cite{Darm98}. This MI gain
does not depend on the wave number q of the carrier wave contrary
to the discrete case.

Based on the analytical results, the modulational instabilities of
carrier wave with wave number $q$ modulated by small oscillation
with wave number $Q$ occurs when the right hand side of dispersion
equation (\ref{eq25}) is negative. In order to check our
analytical results we have performed numerical simulations of
equation (\ref{dcq}) using fourth order Runge-Kutta scheme with a
time step chosen to preserve our conserved quantities which are
the total energy of the system and the number of atoms to accuracy
more than 10$^{-4}$. The numerical simulations have been performed
for a chain of $N=180$ units, with periodical boundary conditions
so that the wave numbers $q$ and $Q$ satisfy the relations $q=2\pi
m/N$ and $Q=2\pi M/N$, where $m$ and $M$ are integers. The other
parameters have been chosen to be $k=0.1$, $\lambda=1$ with
$\gamma$ varying, the amplitude of the modulation wave has been
chosen small compared to the amplitude of the carrier wave as
$\nu_0=0.5$, $\delta \nu_0=0.005$. The stability and instability
regions as it was predicted by the analytical results has been
checked for different points , starting with point labelled (1) in
Fig.1 that correspond to zero quintic term with $q=Q=\pi/3$ and
$q=\pi/3$, $Q=2\pi/3$. Our expectation of modulational instability
has been confirmed as it is shown in Fig. 3 and independently from
$q$ and $Q$.
\begin{figure}[here]
\includegraphics[height=8cm, width=8cm,angle=-90]{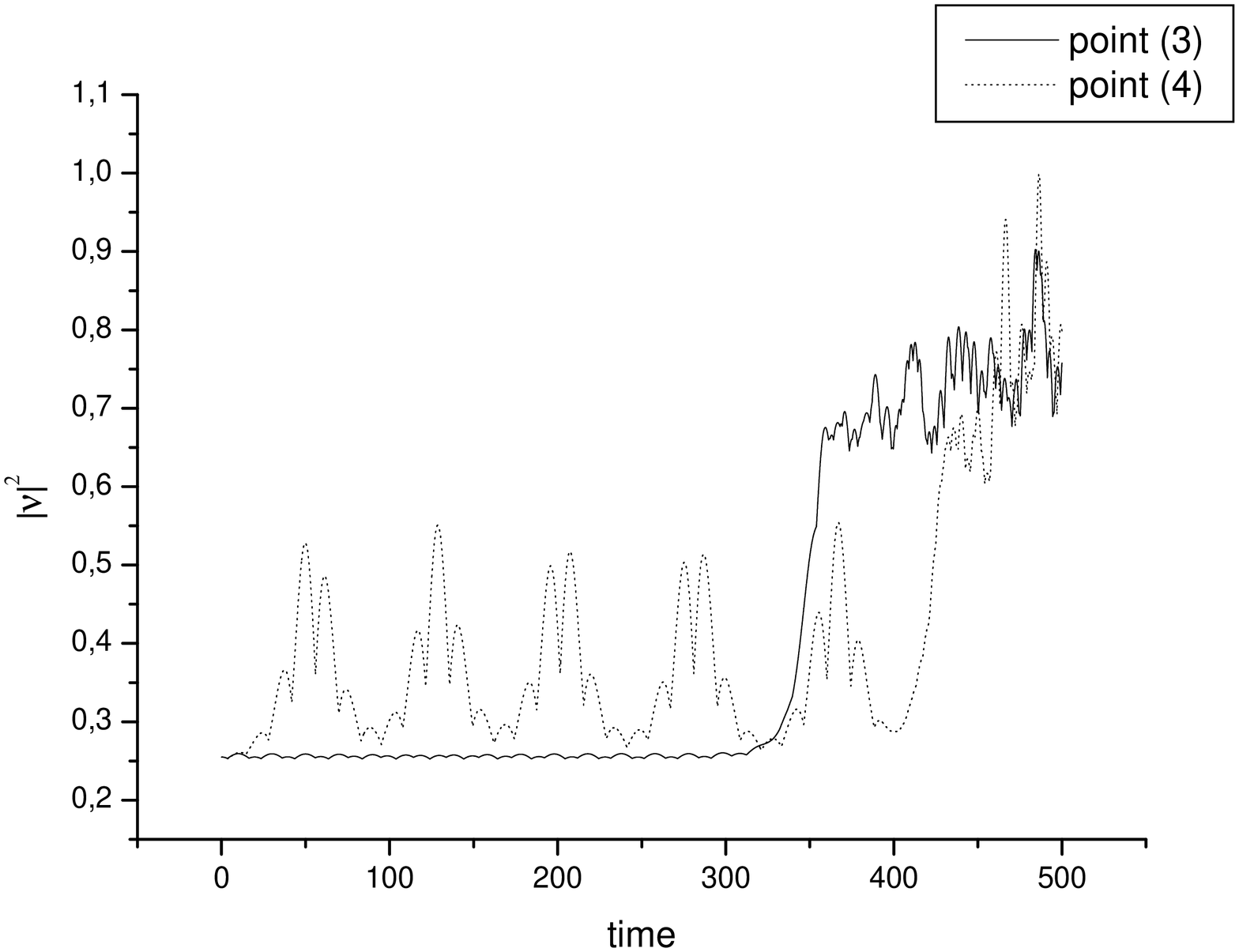}
\caption{Modulational instability arises at different times at
point labelled (3) started immediately and at the point labelled
(4) the solution is stable for a long time before the occurrence
of MI. }
\end{figure}

 The role of
discreteness investigated within the conditional instability
region of Fig. 1 with quintic parameter $\gamma=-1.2$, where we
choose two points: the point labelled (3) with $q=\pi/9$,
$Q=2\pi/3$ and the point labelled (4) with $q=\pi/3$, $Q=2\pi/3$
from Fig.2 that correspond the same point labelled (2) in Fig. 1.
The numerical results demonstrate the dependence of stability on
$q$ and $Q$ as it is shown in Fig. 4.

\begin{figure}[here]
\includegraphics[height=8cm, width=8cm,angle=-90]{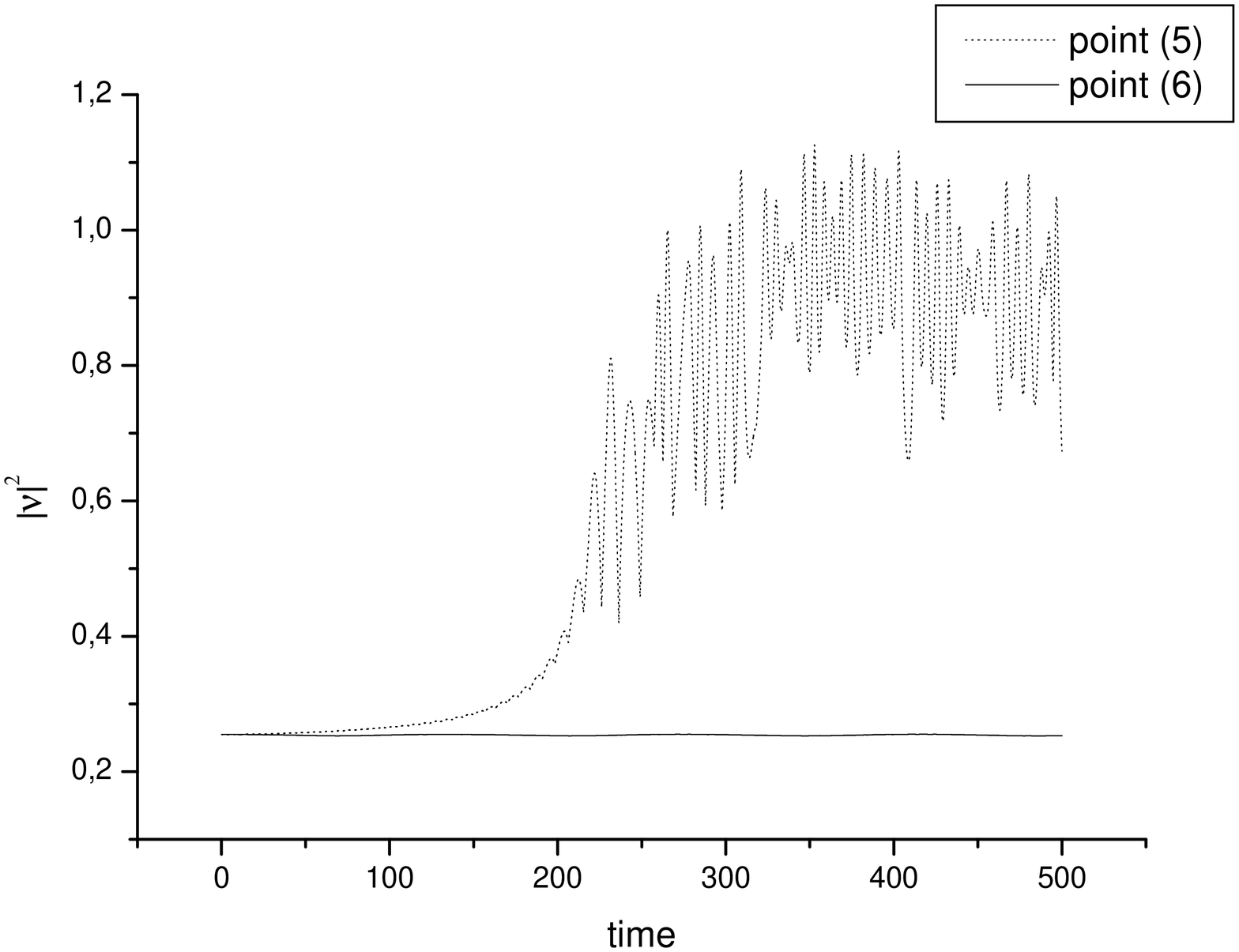}
\caption{ Modulational instability of the plane wave solution
occurs and do not occurs at the same point in plane $(q, Q)$ with
different $\gamma > \gamma_c$ $\gamma < \gamma_c$ }
\end{figure}

\section{Intrinsic localized modes(discrete
breathers)} In this section we study the strongly localized
solutions, which are standing and occupying few sites. In the case
when the coupling between sites are weak the analytical approach
can be used to find the discrete soliton (breather) solutions.

We used Page approach \cite{Page} in order to examine the
existence of the most known localized modes: even, twisted and odd
modes(see Fig. 5). Twisted and even modes has been combined to one
mode where both can be described by one formula as far as the
evolution of one can be other after varying the quintic
nonlinearity term.

\begin{figure}[here]
\includegraphics[height=10cm,width=5cm]{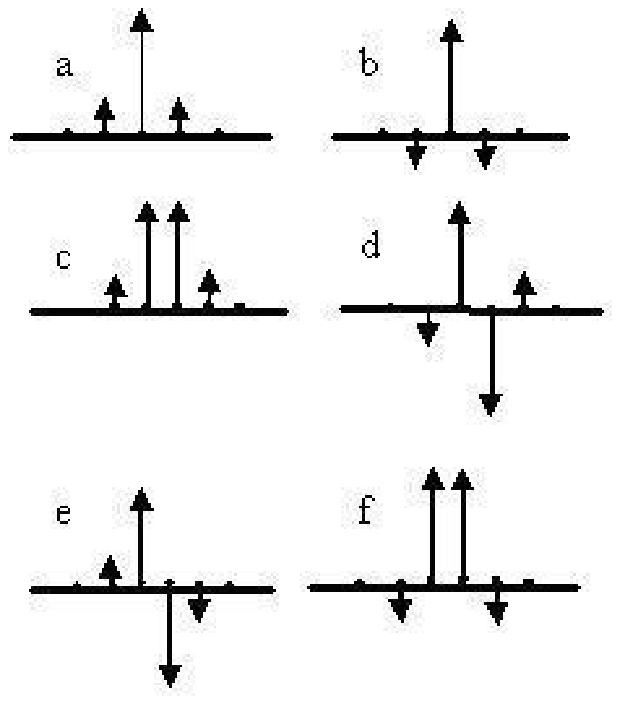}
\caption{Schematic representation of different kind of odd modes
(a unstaggered and b staggered), even modes (c unstaggered and d
staggered)and twisted modes (e unstaggered and f staggered).
 }
\end{figure}

By considering   solution for equation (\ref{dcq}) as
$v_n(t)=\nu_n  {\rm e}^{(i\omega t)}$, where  $\nu_n$ represents
the respective amplitudes of a localized mode one can transform
equation (\ref{dcq}) to a system of algebraic equations depends on
the topology of the localized mode, hence we derive the conditions
of existence and then with linearization we derive the conditions
of stability for each mode as follows.\\

{\bf Even mode}\\

The even/twisted mode defined by its amplitudes as
$v_n=\nu_0\left(\cdots,0,\alpha,1,s,s\alpha,0,\cdots\right)$ ,
where ,  $n=\pm1,\pm2,\cdots$,  as we are concerned with the
symmetric modes only $s=\pm1$. Varying the signs of $s$ and
$\alpha$ produces  the different topologies of even/twisted modes
showed in Fig. 5. Taking in consideration the requirements for
strongly localized modes which are $|\alpha|\ll1$ and
$\alpha_n\approx0$ for $n>2$
 we derive the dispersion relation and the formula of
the secondary
 amplitude respectively as
\begin{eqnarray}
\omega&\equiv&
\omega_{ET}=k(\alpha+s)+\nu_0^2(\lambda+\gamma\nu_0^2)
\label{eq2},
\\
\alpha&=&\frac{k}{\nu_0^2(\lambda+\gamma\nu_0^2)} \label{eq3}.
\end{eqnarray}

Hence a set of conditions for existence of strongly localized
modes can be derived as it is in table 1.
\begin{table}[h]
\begin{center}
\begin{tabular}{|l|c|}
\hline
Mode & Existence conditions\\
\hline
Odd (a) & $\lambda>-\gamma\,\nu_0^2$\\
\hline
Odd (b) & $\lambda<-\gamma\,\nu_0^2$\\
\hline
Even (c) and Twisted (e) & $\lambda>-\gamma\,\nu_0^2$\\
\hline
Even (d) and Twisted (f)& $\lambda<-\gamma\,\nu_0^2$\\
\hline
\end{tabular}
\caption{ Conditions for existences of different kinds of modes
shown in Fig. 5 and for positive coupling.}
\end{center}
\label{tab1}
\end{table}

To study the stability of these modes we followed   the approach
developed in  \cite{Darm98} by using a linear analysis, where we
impose a perturbation $\delta_n(t)$ on each non zero excitation
amplitude, hence the mode's amplitude can be written now as
$$\nu_n=\nu_0(\cdots,0,r\alpha+\delta_{-2},1+\delta_{-1},
s+\delta_1,\alpha+\delta_2,0,\cdots),$$ inserting it in
(\ref{dcq}) and with  subsequent linearization we got eight-order
system of equations. The change of variables as
$\delta^{\pm}_j=\delta_{+j}\pm\delta_{-j}$ ($j$=1,2) reduces the
system to two independent four-order equations systems. Separating
the real and imaginary parts of the perturbation
$\delta^{\pm}_j=\delta^{\pm}_{jr}+i\,\delta^{\pm}_{ji}$
   and introducing the scaled time $\tau=\omega_c\,t$,
we obtain

\begin{widetext}
\begin{eqnarray}
\frac{d \delta^{\pm}}{d \tau}= \left(
\begin{array}{cccc}
0& (s-p)\alpha & 0 &\alpha\\
2-(3s-p)r\alpha+\displaystyle\frac{2\gamma\nu_0^2(1-r\alpha)}{(\lambda
+\gamma\nu_0^2)}
&0&\alpha&0\\
0& -\alpha&0&1\\
\alpha&0&-1&0\\
\end{array}
\right)\delta^{\pm}, \label{eq4}
\end{eqnarray}
\end{widetext}
where
$\delta^{\pm}=(\delta^{\pm}_{1r},\delta^{\pm}_{1i},\delta^{\pm}_{2r},
\delta^{\pm}_{2i})$ and  $p=\pm1$ stands for the symmetric
$\delta^+_j$ and antisymmetric $\delta^-_j$ perturbation
respectively. If we introduce $\delta^{\mp}\propto {\rm
exp}(g\tau)$ then the eigenvalues $g$ of (\ref{eq4}) are given by
the following equation

\begin{widetext}
\begin{eqnarray}
&&g^4+\left[ 1+ \left(
\frac{2\gamma\nu_0^2(p-s)}{(\lambda+\gamma\nu_0^2)}+2(3-2ps)
\right)\alpha^2+
2(p-s)\left(1+\frac{\gamma\nu_0^2(p-s)}{\lambda+\gamma\nu_0^2})
\right) \alpha \right]g^2
\nonumber\\
&&+ 2(p-s)\left( 1+\frac{\gamma\nu_0^2}{\lambda+\gamma\nu_0^2}
\right)\alpha+ \left(
2(3-2ps)+\frac{2\gamma\nu_0^2(1-p+s)}{\lambda+\gamma\nu_0^2}
\right)\alpha^2
\nonumber\\
&&+\left(
2(p-2s)-\frac{2\gamma\nu_0^2(1-p+s)}{\lambda+\gamma\nu_0^2}
\right)\alpha^3+\alpha^4=0. \label{eq5}
\end{eqnarray}
\end{widetext}
Hence similarly to the case without quintic nonlinearity
\cite{Darm98}, when the symmetry of the mode coincides with the
symmetry of perturbation that is $p=s$  the localized modes will
be stable without any conditions on the nonlinearity parameters.
In the case where the mode and the perturbation have different
symmetries that  $p=-s$ then there is a possibility for the mode
to be unstable with the instability gain
\begin{widetext}
\begin{eqnarray}
g=-2\sqrt{1+\frac{\gamma \nu_0^2}{\gamma
\nu_0^2+\lambda}}\sqrt{s\alpha}\left(1 -
\left(\frac{5s}{4}+\frac{\gamma \nu_0^2(1-2s)}{2(2\gamma
\nu_0^2+\gamma)} \right)\alpha \right), \label{eq6}
\end{eqnarray}
\end{widetext}


In contrast to the case of cubic nonlinearity only, where the mode
is stable only when  $\alpha s<0$, which means only twisted modes
( staggered and unstaggered ) are stable, it is clear in the
cubic-quintic case that  the first coefficient of (\ref{eq6})
leads also to the possibility of  stability of even modes
(staggered and unstaggered) $\alpha s>0$. However taking the
existence consideration only the unstaggered even mode is stable.
The relationships between nonlinearities that control the
stability is given in table 2.
\begin{table}[h]
\begin{center}
\begin{tabular}{|l|c|}
\hline
Mode & Stability Conditions\\
\hline
Even (c \& d) & $\lambda<-2\gamma\,\nu_0^2$\\
\hline
Twisted (e \& f) & $\lambda>-2\gamma\,\nu_0^2$\\
\hline
\end{tabular}
\caption{Conditions for stability of even/twisted modes showed in
Fig. 1}
\end{center}
\end{table}
In the numerical simulations, first we checked the stability of
the even modes when the symmetry of the perturbation is the same
as the symmetry of the mode. We found that the numerical results
confirm the analytical predictions and the modes are stable
always. For the case when the perturbation and the mode have
opposite symmetry we checked the validity of analytically derived
stability regions for the modes.

The most important result here is  that  the predicted stable even
unstaggered mode, which is not possible in the case with cubic
nonlinearity only, see \cite{Darm98} and references therein, has
been demonstrated numerically see Fig. 6 for $\gamma = -0.6$ which
is less than $\gamma= \lambda/2{v_{0}}^2 =-0.5$. However when
$\gamma=-0.4$ which is greater than $\gamma= \lambda/2{v_{0}}^2
=-0.5$, the mode is unstable and it is transformed to the odd mode
after some time as it is shown in Fig. 7. The even twisted modes
are shown analytically to be stable within two different regions
and this result has been checked numerically, which is in
agreement with \cite{Darm98} for the case of cubic nonlinearity only.\\

\begin{figure}[here]
\includegraphics[height=8cm, width=6cm,angle=-90]{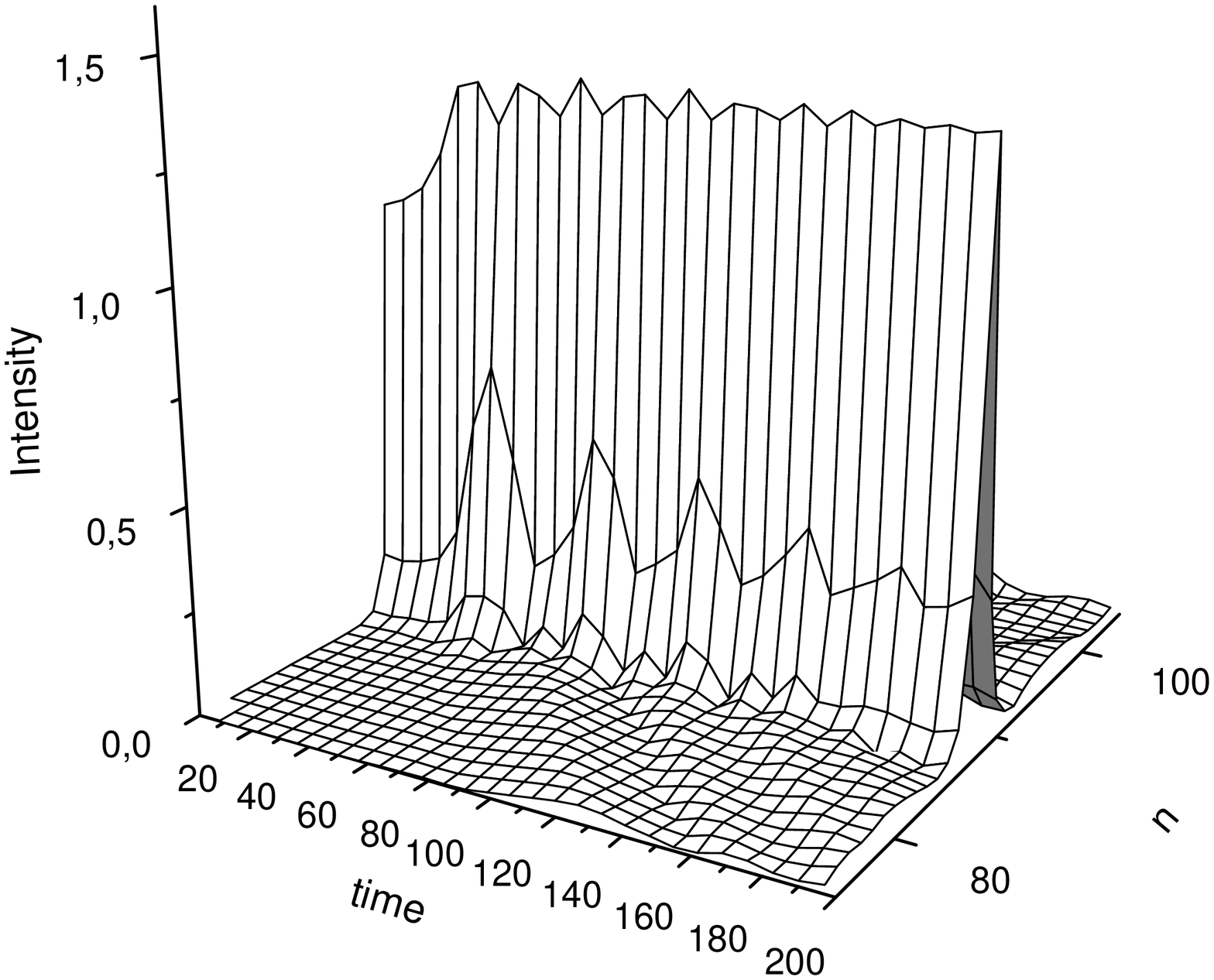}
\caption{Evolution of perturbed unstaggered even mode (c) $\lambda
=1, v_{0}=1, \gamma=-0.6$. }
\end{figure}

\begin{figure}[here]
\includegraphics[height=8cm, width=6cm,angle=-90]{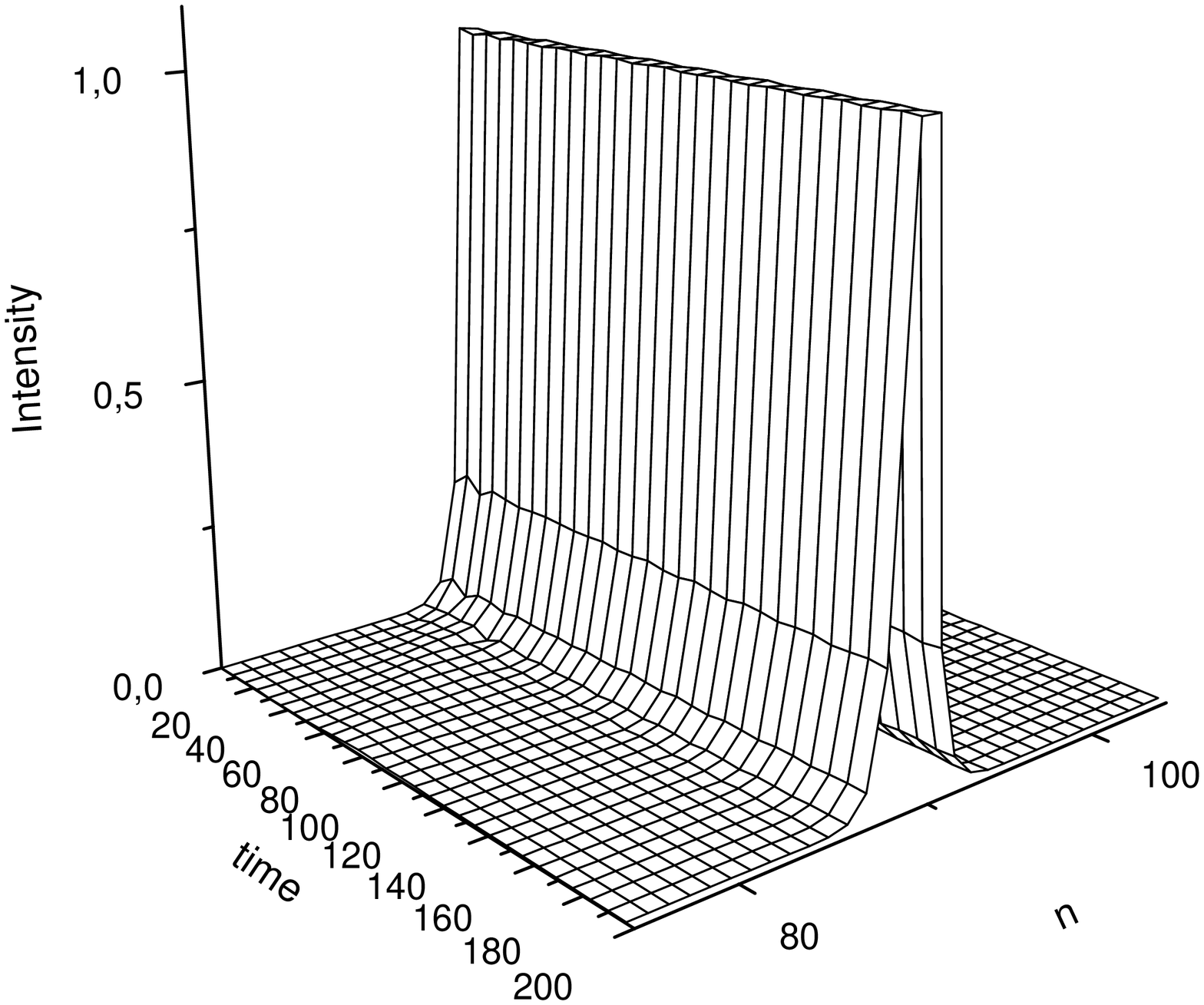}
\caption{Evolution of perturbed unstaggered even mode (c) $\lambda
=1, v_{0}=1, \gamma=-0.4$. }
\end{figure}

{\bf Odd mode}\\

The odd mode is defined by its amplitude as
$\nu_n=\nu_0(\cdots,0,\beta,1,s\beta,0,\cdots)$, where
$n=0,\pm1,\pm2,\cdots$  and $s=\pm1$ with the requirements for
strong localized modes are $\beta\ll1$ and $\beta_n\approx 0$ for
$n>1$ and after performing similar algebraic operations to those
done for even/twisted mode  we obtain the dispersion relation and
the formula of the secondary amplitude respectively as .
\begin{eqnarray}
\omega&\equiv&
\omega_o=k^2\frac{1+s}{\nu_0^2(\lambda+\gamma\nu_0^2)}+\lambda\nu_0^2+\gamma\nu_0^4,
\label{eq7}
\\
\beta&=&\frac{k}{\nu_0^2(\lambda+\gamma\nu_0^2)}, \label{eq8}
\end{eqnarray}
which leads to a set of conditions for the existence of this mode
as it is shown previously in table 1. To study the stability of
this mode we insert perturbation  $\delta_{n}(t)$ on each non zero
excitation amplitude as
$$\nu_n=\nu_0(\cdots,0,\beta+\delta_{-1},1+\delta_0,s\beta+\delta_1,0,\cdots),$$
then substitute it in equation (\ref{dcq}) and perform
linearization procedures, as we did for the even/twisted mode, we
got  a six-order system of equations, again a change of variables
as
 $\delta_1^+=\delta_{+1}+\delta_{-1}$ reduces the system
to four-order equations system.
 Separating the real and imaginary parts of the
perturbation $\delta^+_1=\delta_{+1r}+i\,\delta_{1i}$ and
$\delta_0=\delta_{0r}+i\,\delta_{0i}$, then introducing the scaled
time  $\tau=\omega_c t$ we obtain
 \begin{eqnarray}\label{eq35}
\frac{d\delta^{\rm od}}{d\tau}= \left(
\begin{array}{cccc}
0&0&0&-\beta\\
2+\displaystyle\frac{\gamma\nu_0^2}{\lambda+\gamma\nu_0^2}&0&\beta&0\\
0&-2\beta&0&1\\
2\beta&0&-1&0
\end{array}
\right)\delta^{\rm od},
\end{eqnarray}

where  $\delta^{\rm od}=(\delta_{0r}^{\rm od},\delta_{0i}^{\rm
od}, \delta_{1r}^{\rm od},\delta_{1i}^{\rm od})$ . If we introduce
$\delta^{\mp}\propto {\rm exp}(g\tau)$ then the eigenvalues $g$ of
(\ref{eq35}) are given by the following equation
\begin{eqnarray}
g^4+\left[1+4s\beta\right]g^2+4\left[1+\frac{\gamma\nu_0^2}{\lambda+
\gamma\nu_0^2}+\beta^2\right]\beta^2=0, \label{eq36}
\end{eqnarray}
which have four roots and  the one that  control the stability is
\begin{equation}
g=\frac{1}{2}\sqrt{-2-8\beta^2+2\sqrt{1-24\beta^2+
\frac{16\lambda\beta^2}{\lambda\nu_0^2+\lambda}}}, \label{eq37}
\end{equation}
which shows that the instability gain Re($g$) will be a nonzero
only if $|\beta|>\beta_c$ with
$\beta_c=\displaystyle\pm\frac{1}{\sqrt{8}}\sqrt{\frac{\lambda+
\gamma\nu_0^2}{\lambda+3\gamma\nu_0^2}}$. For $\gamma =
\gamma_c=-\lambda/3\nu_0^2$, $\beta_c$ will be undefined, but we
can conclude from equation (\ref{eq36}) that the odd mode will be
stable whatever the value of $\beta$. It is important to mention
here that the value of  $\beta_c$ is depend on the interplay
between the cubic and quintic nonlinearities and satisfies the
condition for strongly localized modes $\beta\ll1$ for a range of
values of these nonlinearities, in contrast to the case of the
CDNLSE discussed in \cite{Darm98} where $\beta_c = 1/\sqrt{8}$
which showed that the strongly localized mode is always stable
because $\beta < \beta_c$ always. Furthermore for the stability of
localized odd mode in CQDNLSE the condition
$\gamma>\gamma_c=-\lambda/3\nu_0^2$ is necessary for focusing
cubic nonlinearity and defocusing quintic nonlinearity. Finally we
should mention here that for both even/twisted modes and odd modes
cases, if we substitute $\gamma=0$ in our results, we get back to
the cubic case and hence we reproduce the results of
\cite{Darm98}. Numerical results for the case of odd modes
presented in Figs. 8-9, confirmed the new predicted result that is
the quintic nonlinearity term can lead to an unstable staggered
odd mode when $-2<\gamma<-1.5$, this range is chosen to keep the
mode strongly localized means the second amplitude $0.1\ \le \beta
\le 0.2$. However the unstaggered odd mode is stable always as it
is shown in Fig. 10 in  agreement with the results for the case of
cubic nonlinearity only \cite{Darm98}.

\begin{figure}[here]
\includegraphics[height=8cm, width=6cm,angle=-90]{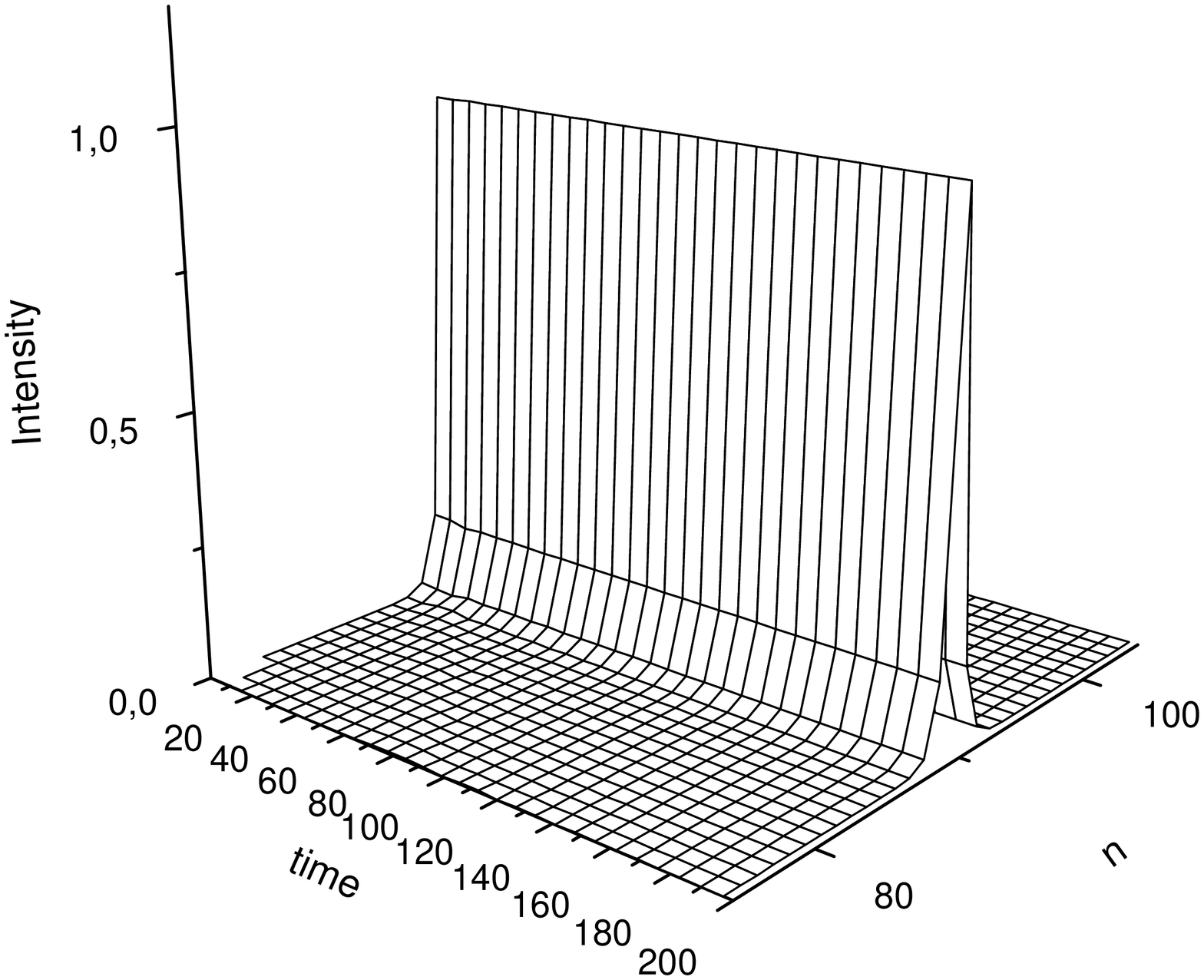}
\caption{Evolution perturbed of staggered odd mode (b) $\lambda
=1, v_{0}=1, \gamma=-1.7, \beta_c=0.140 < \beta=0.167$. }
\end{figure}

\begin{figure}[here]
\includegraphics[height=8cm, width=6cm,angle=-90]{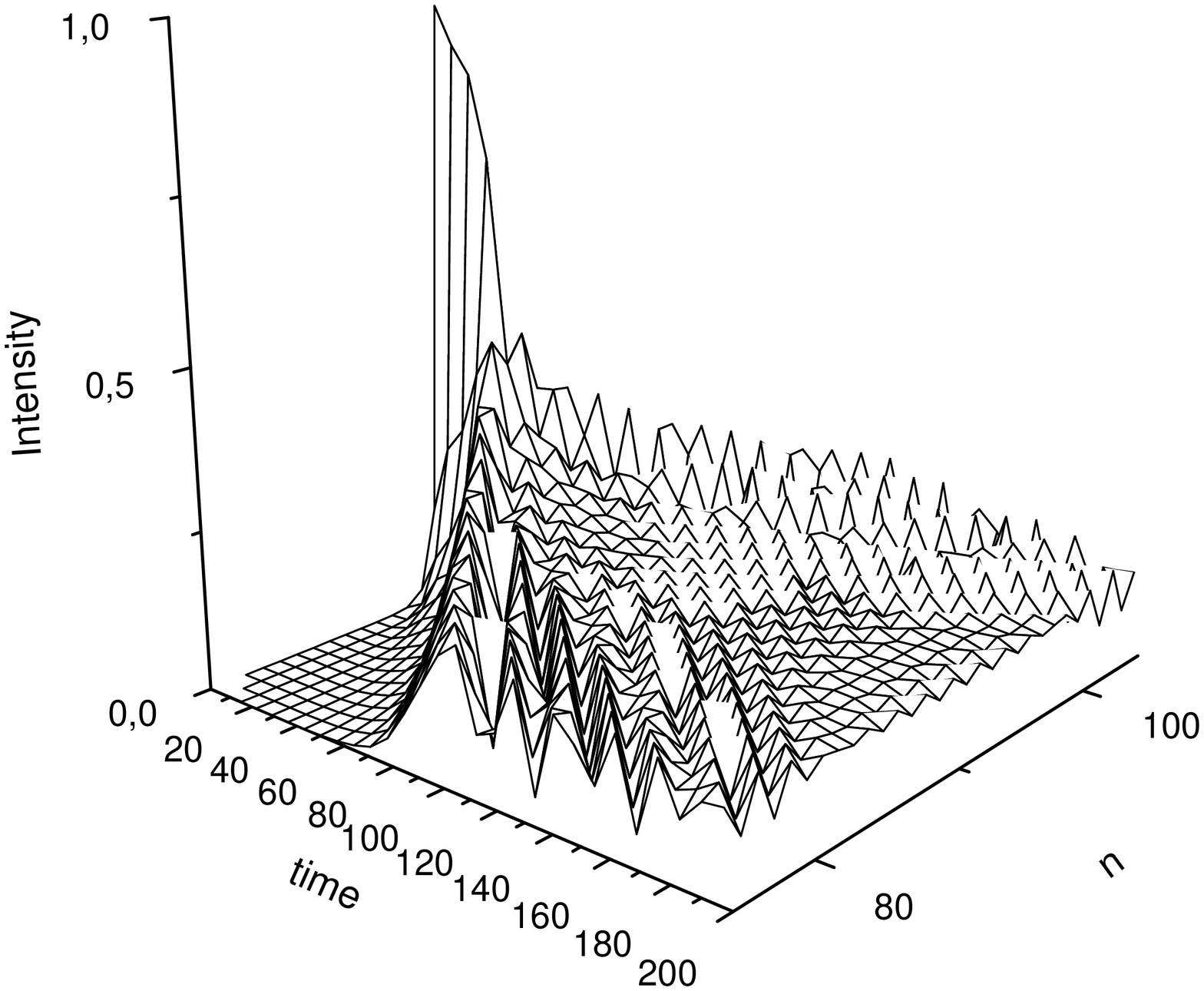}
\caption{Evolution of perturbed staggered odd mode (b) $\lambda
=1, v_{0}=1, \gamma=-1.6, \beta_c=0.146 > \beta=0.143$. }
\end{figure}

\begin{figure}[here]
\includegraphics[height=8cm, width=6cm,angle=-90]{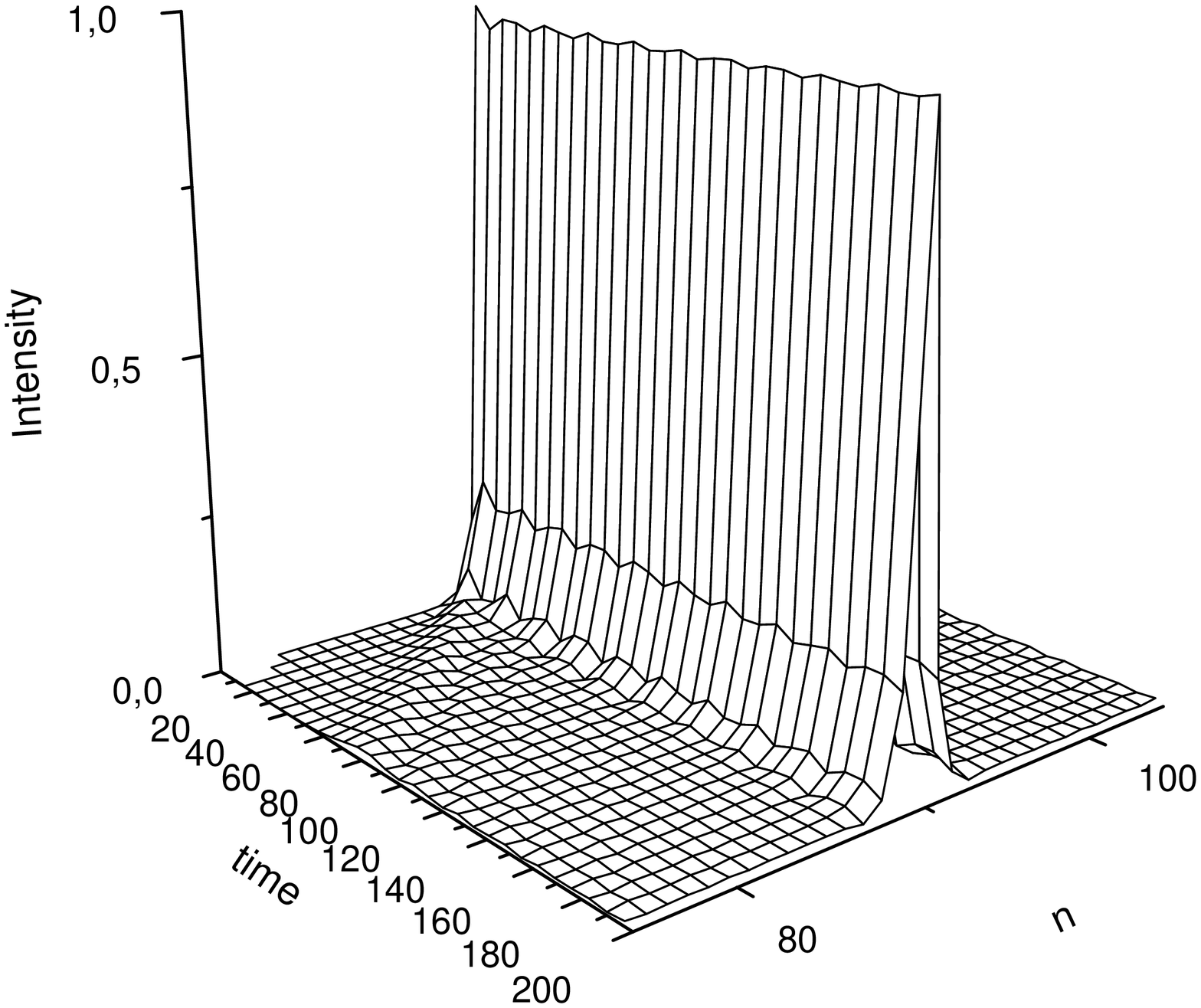}
\caption{Evolution of perturbed unstaggered odd mode (a) $\lambda
=1, v_0=1, \gamma=-0.3$.
 }
\end{figure}

\section{Conclusion}
In conclusion we have investigated the modulational instabilities
and discrete breathers in BEC with two- and three-body
interactions in the optical lattice. Following to the procedure
suggested in \cite{Alfimov} we derive the lattice nonlinear model
which describe the GP equation with periodic potential. This
approximation is reasonable for the atomic wave function is
approximated by a single Wannier function - so called a Wannier
soliton. The modulational instability in a lattice is the process
leading to the generation of discrete breathers in the optical
lattice. We find the regions of instabilities of nonlinear plane
wave and conform analytical predictions by the direct numerical
simulations of the CQDNLS equation. The lattice dispersion vary
sign in dependence of the quasi-momentum value in the Brilloin
zone. Thus we can control the regions of instability even for the
fixed signs of two and three body interactions. This circumstance
extend the possibility for creation of the localized modes in BEC
in optical potential. We find the expressions for the amplitudes
and frequencies of different strongly localized modes in optical
lattice such as even, odd, twisted stuggered and unstaggered
modes. We analyze the stability of these modes and found the
stability regions which are necessary for the search of these
modes in the experiments with BEC in optical lattice. For the
further investigations will be interesting to consider the regimes
beyond of the tight-binding approximation(the deep optical lattice
limit) by accounting the higher order terms in the Wannier
expansion. In particular it concerns the extensions of CQDNLSE
involving the discrete nonlinear models with nonlocal
nonlinearities\cite{TS2,Alfimov,KevPel,Khare}.

\end{document}